\documentclass[pre,twocolumn]{revtex4-1}

\usepackage{verbatim}   
\usepackage{subfigure}  
\usepackage[colorlinks=true, citecolor=blue, urlcolor=blue,,linkcolor=blue]{hyperref}   
\usepackage[utf8]{inputenc}
\usepackage[english]{babel}
\usepackage{amsmath}
\usepackage{amsfonts}
\usepackage{amssymb}
\usepackage{graphicx}
\usepackage[export]{adjustbox}
\usepackage{color} 
\setcitestyle{numbers}
\usepackage[normalem]{ulem}

\begin{document}

\title{Simulation of surfactant transport during the rheological relaxation of two-dimensional dry foams}
\thanks{{\em Preprint accepted by Physical Review E}}
\author{F. Zaccagnino}
\affiliation {Department of Mathematics, Aberystwyth University, SY23 3BZ, UK}
\author{A. Audebert }
\affiliation {STLO, UMR1253, INRA, Agrocampus Ouest, 35000, Rennes, France}
\author{S.J. Cox}
\affiliation {Department of Mathematics, Aberystwyth University, SY23 3BZ,UK}

\begin{abstract}
   We describe a numerical model to predict the rheology of two-dimensional dry foams.
The model accurately describes soap film curvature, viscous friction with the walls, and includes the transport of surfactant within the films and across the vertices where films meet. It accommodates the changes in foam topology that occur when a foam flows and, in particular, accurately represents the relaxation of the foam following a topological change. The model is validated against experimental data, allowing the prediction of elastic and viscous parameters associated with different surfactant solutions.
\end{abstract}

\maketitle

\section{Introduction}

An aqueous foam is a collection of gas bubbles in a surfactant-laden liquid. They are used widely, for example in ore separation, fire-fighting and interventional medicine~\cite{cantat2013foams}, and therefore understanding the flow properties of this yield-stress fluid is important. The stability imparted to the thin films by the surfactant is tenuous, and as a foam flows the local concentration of surfactant may vary, leading to variations in surface tension and possible film collapse. 

A dry foam is a foam characterized by low liquid (or continuous phase) fraction. The local geometry of the soap films, which is described by Plateau's laws~\cite{cantat2013foams}, is determined by mechanical and thermochemical equilibrium conditions, and the Young-Laplace law describes the fundamental relationship between the curvature of the soap film and the pressure drop across it~\cite{cantat2013foams}.
As a foam flows, the bubbles move past one another, leading to the film-scale topological processes of foam rearrangement known as T1s. 

A T1 can be thought of as an evolution from an initial non-equilibrium configuration toward a final equilibrium configuration. 

\begin{figure}[!htbp]
	\centering
    \includegraphics[width=0.35\textwidth]{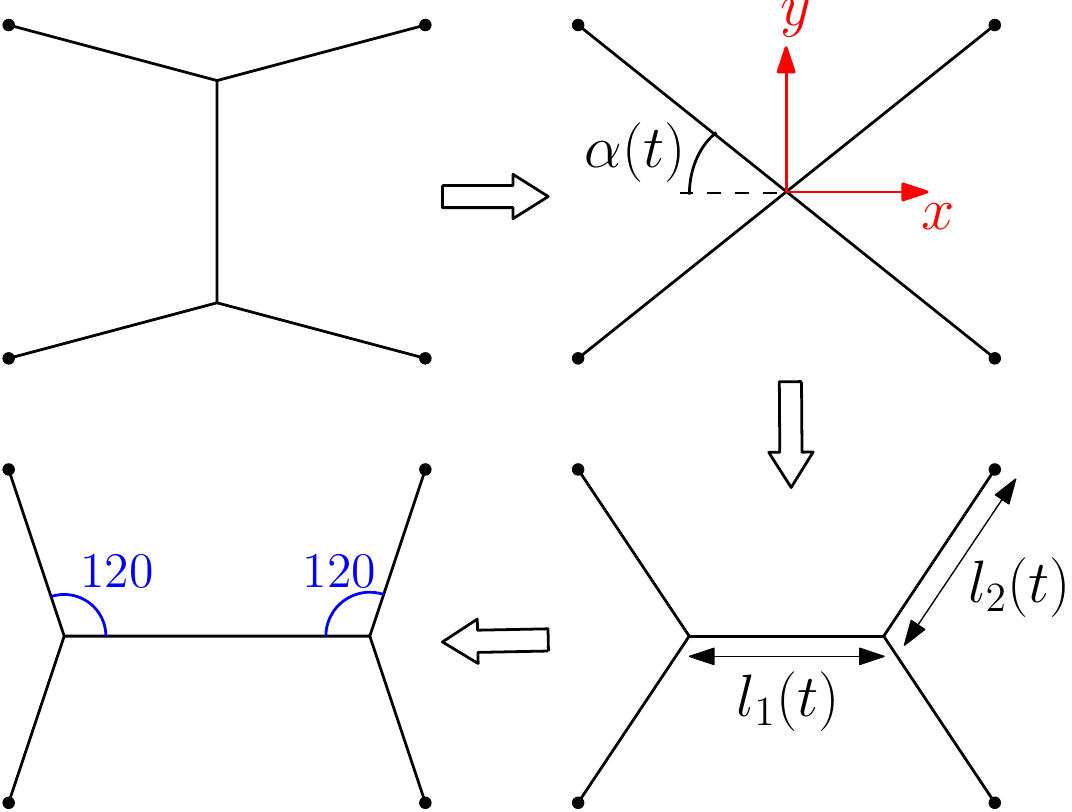}
\caption{Topological T1 process: evolution from the starting configuration to the final one in which the surface tension forces are balanced and $ \alpha = {60}$ degrees.}
\label{fig:T1}
\end{figure}

It is quite straightforward to realize an experiment to study a single T1. Consider the evolution of five soap films between two flat plates which are connected by four pins~\cite{cantat2013foams}, as shown in Fig. \ref{fig:T1}. The initial configuration collapses, due to the shrinking of the central film, into an unstable configuration in which four films meet at a point. Instantaneously a new film is created. The new film will stretch, while the lateral films shrink~\cite{cantat2013foams}, until equilibrium is reached. The films are subjected to the same air pressure on each side (so that we need not consider the effects of pressure). Nonetheless, during the relaxation of the structure the films are not straight: the lateral film, denoted $ l_2 $ in Fig. \ref{fig:T1}, exhibits a smooth curvature. This is due to the effects of drag with the flat plates bounding the films. 
Here we focus on this single T1: we investigate the evolution of the new film generated during the T1 in relation to the surface rheological properties of the film. In particular we look at the influence of the viscoelastic parameters on the film evolution toward the final equilibrium configuration. 

We study the thin layer at the interface where the surfactant molecules create an ordered array. We focus purely on the interfacial phenomena and we neglect all the diffusion and absorption processes. We therefore consider an insoluble layer, by assuming that on each interface the rate of surface dilatation greatly exceeds the rate of surfactant transport to or from the bulk phase.

There are several quasi-static models that have been developed to simulate foam flow on the basis of energy minimization~\cite{cantat2013foams}. Here, we wish to investigate viscous phenomena, and so we instead choose a dynamical model in which the film shape and motion are determined by a force balance.
The Viscous Froth (VF) model~\cite{kern2004two}, which we present in section {\ref{VF}}, was developed to extend quasi-static models of two-dimensional foam flow to include the viscous drag that is exerted on the soap films by the surfaces bounding the foam. It is adapted particularly to a bubble monolayer in a Hele-Shaw cell (between parallel glass plates), and shows how rate effects influence film shape~\cite{drenckhan2005rheology}.

However, the VF model considers surface tensions to be constant, so we add to the VF model a surfactant transfer (ST) model. As a consequence we are able to describe the variation of surface tension as a function of the concentration of surfactant molecules on each of the films of a foam. The evolution of surfactant concentration $ \Gamma $ and surface tension $ \gamma $ are related by the Langmuir equation of state:
\begin{equation}
\gamma = \gamma_{eq} - E  \ln \frac { \Gamma}{\Gamma_{eq}} ,
\label{eq:Langmuir}
\end{equation} 
where $ E $ is the Gibbs elasticity, while $ \gamma_{eq} $ and $ \Gamma_{eq} $ are the values of the surface tension and the surfactant concentration at equilibrium. We also introduce an additional viscous term which accounts for the motion of surfactants due to gradients in surface tension.

Our model extends the work of Durand and Stone~\cite{durand2006relaxation}, in which the dynamics of films after a T1 follows from a force balance (at the vertex) between the stretching and the shrinking films and a surfactant mass balance. The DS model considers the three films $ l_1 $, $ l_2 $ and $ l_3 $, shown in Fig. \ref{fig:s1},  and assumes them to be straight. The monotonic increase of the angle $ \alpha $ between the lateral film, $l_2$, and the $ x $ axis, after a T1, drives the stretching of the newly-created central film, $l_1$. 

The DS model also considers that on the shrinking interfaces the surfactant concentration (and hence surface tension) is constant, and so the equilibrium condition of equal film tensions, when films meet at $ \frac{2 \pi}{3}$, is only realised in the limit of $ E \rightarrow 0 $ (this result is compared with our own, later, in Fig. \ref{fig:U6}). In the force balance, the DS model combines the shear $\mu_s$ and dilatational $k$ viscosities into a single surface dissipative term. They fit experimental data for two surfactants: sodium dodecyl sulfate (SDS) and protein bovine serum albumin (BSA), in order to estimate the characteristic Gibbs elasticity and the sum of shear and dilatational viscosities for these foam films, in good agreement with values present in the literature~\cite{liu2003effects,sarker1999restoration}.
  
\begin{figure}[!htbp]
	\centering
	\includegraphics[width=0.6\linewidth]{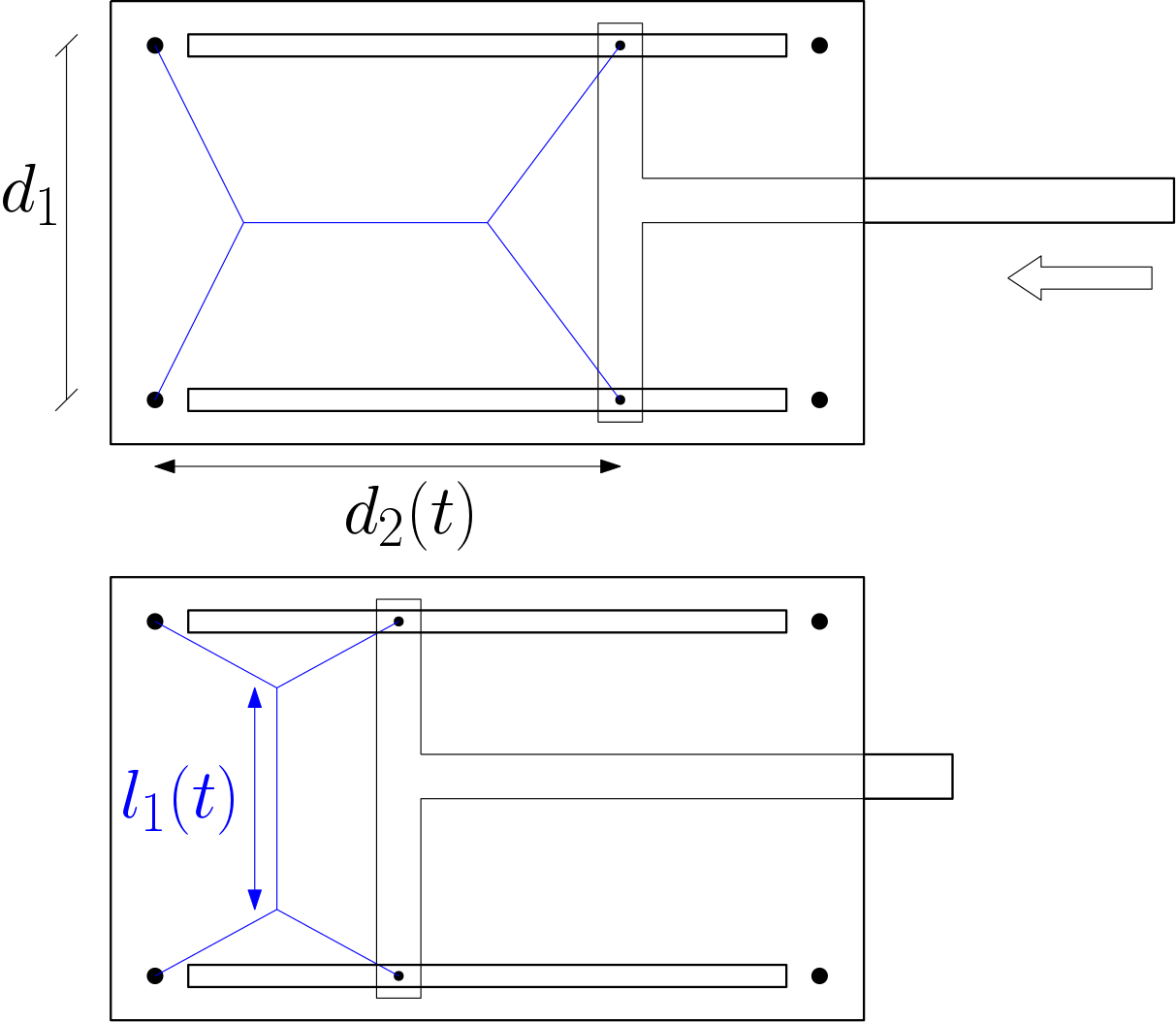}
	\caption{Device used for the experimental measurements.}
	\label{fig:m1}
\end{figure}

Relaxing these rather strict assumptions, we will allow film curvature and variations in surface tension on all films. We will validate our new VF+ST model by fitting its predictions to experimental data for the evolution of the length of the newly created film after a T1 in systems containing different surfactants. 
Extending the VF model, which generates an estimate of the viscous drag coefficient, our VF+ST model predicts two additional parameters, the Gibbs elasticity and the surface viscosity, for each surfactant mixture. Although at very short times the VF model is able to fit the data for anionic surfactants, at long times the introduction of the additional viscous factors within the VF+ST model is crucial in order to fit data for both anionic surfactants and proteins.

\section{The Model}

\subsection{Surfactant transport model}

\begin{figure}[!htbp]
	\centering
	\includegraphics[width=0.4\textwidth]{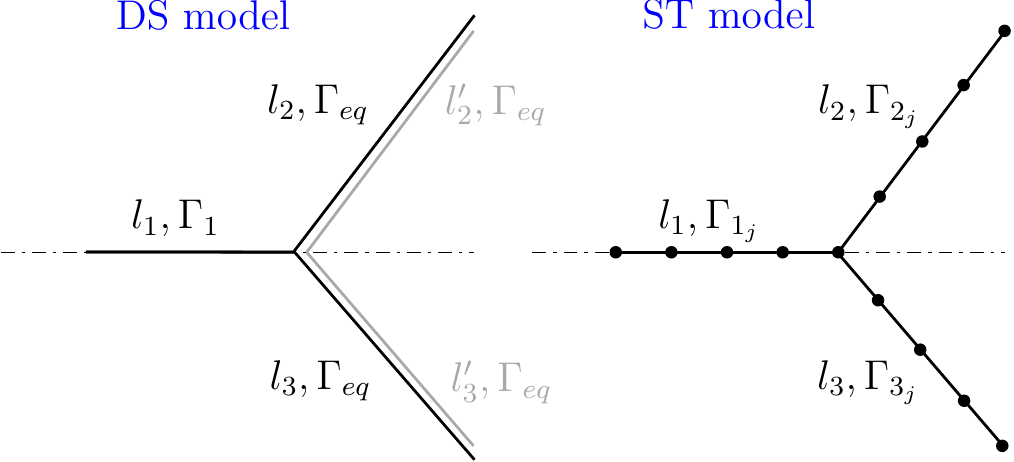}
	\caption{On the left, the film configuration considered in DS model, $l_1$ $l_2$ and $l_3$. The shrinking films, $l_2$ and $l_3$, have constant surfactant concentration, $ \Gamma_{eq} $. The dot-dashed line represents the axis of symmetry. On the right, the configuration considered with the ST model.  Each film $l_i$ is discretized into several short straight segments $l_{i_j}$, each with a variable surfactant concentration $\Gamma_{i_j}$.}
	\label{fig:s1}
\end{figure}

Here we model the variation of surfactant coverage $ \Gamma $ along the films. To allow film curvature, each film is discretized into a number of short straight segments. We define these segments to meet at {\em points} within the film, and three segments meet at a {\em vertex}. We introduce subroutines to keep the average length of segments uniform, by subdividing and removing segments when they become too large or short respectively, ensuring that the number of surfactant molecules is conserved. In particular we check at each time that the segment length $l_j$ is always $l_{min} \leq l_j \leq l_{max}$ where $l_{min}$ is the length at which a T1 may occur (if the short segment directly connects two vertices). In this way we allow compression and stretching of films and, by choosing an appropriate time-step, ensure the stability of the numerical calculation.

We calculate the variation of surface tension $ \gamma $ through the equation of state, Eq.~(\ref{eq:Langmuir}), applied to each segment. We define a ``convection equation", similar to the Marangoni effect~\cite{edwards1991interfacial}, to describe the movement of surfactant molecules along each film as a consequence of gradients of surface tension. This takes the form of a rule for tangential motion of the points, which connect pairs of film segments, along the film, based on the difference in tension between the two segments. As a consequence, the segments contract when they have high surface tension, leading to an effective flow of surfactant within the film. This flow is subject to a viscous drag, and so our model balances the gradient of surface tension along each film with the tangential component of the surface velocity, $ v_t $, multiplied by a factor $ \mu $, which we think of as a surfactant drag coefficient:
\begin{equation}
\mu v_t(s) -\frac {\partial{\gamma}} {\partial{s}} = 0 ,
\label{eq:Marangoni}
\end{equation}
where $ s $ is the curvilinear coordinate along the film. 
The gradient of surface tension is known at each point, so  we are able to apply this equation there, rather than considering only the balance at the vertex, as is done in the DS model. 

\begin{figure}[t]
	\centering
    \includegraphics[width=0.2\textwidth]{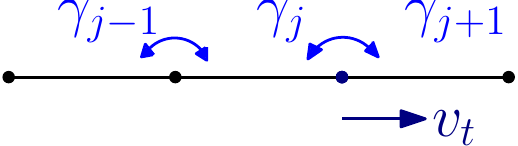}
\caption{A film is discretized into several segments. Each segment has an initial number of molecules and surfactant concentration. Through Eqs.~(\ref{eq:Langmuir}) and (\ref{eq:Marangoni}) we calculate the surface tension $\gamma_j$ of each segment and the tangential velocity component $v_t$ of each point.}
\label{fig:film}
\end{figure}

The surfactant balance is imposed numerically: we assume a starting value for the surfactant concentration, we calculate the number of surfactant molecules $N_{mol}$ per segment, through the expression $ \Gamma_{j} = \frac{N_{mol}}{l_{j}} $, and then we update the surfactant concentration of each segment, according to its deformation. 
We are therefore able to calculate the surface tension for each segment by means of Eq.~(\ref{eq:Langmuir}), and then applying Eq.~(\ref{eq:Marangoni}) gives the tangential velocity component for each point, see Fig.~\ref{fig:film}.

\subsubsection{Is diffusion of surfactant important?}

Within a real foam film, surfactant molecules move in several ways. As we note above, we assume that surfactant transfer between the surface layer and the bulk is slow compared to the motion within the surface layer. Within the surface layer itself, surfactant molecules respond to gradients in concentration: this can take the form of a convective motion, limited by viscosity, as in Eq.~(\ref{eq:Marangoni}), or a diffusive flow.

In order to highlight the qualitative behaviour of Eq.~(\ref{eq:Marangoni}), we compare it with the result of solving a diffusion equation. We consider a single straight film of unit length with an initial Gaussian distribution for the concentration of surfactant molecules, centred on the middle of the film, and we compare the standard analytical solution for the diffusion equation with a numerical solution of Eq.~(\ref{eq:Marangoni}).

\begin{figure}
	\centering	
		\subfigure[\label{diff1a}]{\includegraphics[width=.95\linewidth,valign=t]{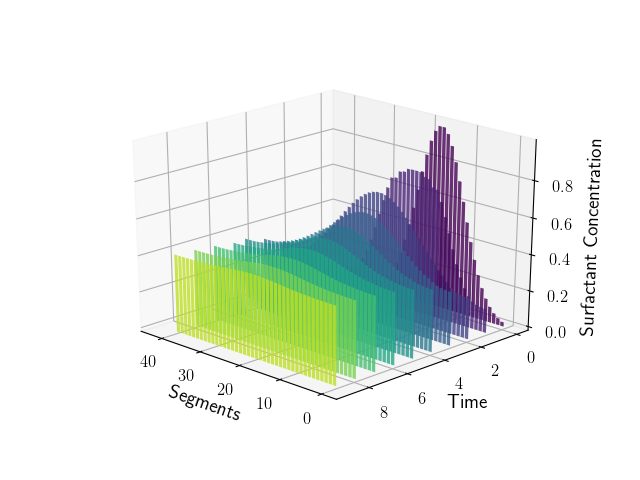}}
		\subfigure[\label{diff1b}]{\includegraphics[width=1.0\linewidth,valign=t]{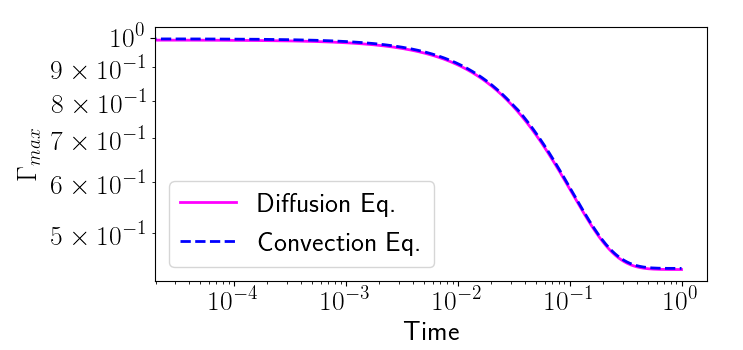}}
	\caption{a) Variation of the concentration of surfactant molecules on a single straight film, starting from a Gaussian distribution, from the solution of Eq.~(\ref{eq:Marangoni}) with $\mu = 1$. We show the variation of surfactant concentration on each segment with time as it evolves towards the final, uniform, distribution. b) Comparison of the maximum value of $\Gamma$ between the numerical solution of Eq.~(\ref{eq:Marangoni}) and the analytic solution of the diffusion equation with unit diffusion coefficient.}
	\label{fig:diff1}
\end{figure}

Fig. \ref{diff1a} shows the variation of surfactant concentration on each segment with time as it evolves towards the final, uniform distribution, calculated through Eq.~(\ref{eq:Marangoni}).
Fig. \ref{diff1b} shows how the two phenomena, plotted in {\em dimensionless} form, are identical. Thus Eq.~(\ref{eq:Marangoni}) is effectively acting as a diffusive process.

However, in {\em dimensional} form, the typical order of magnitude for the diffusion coefficient for surfactant motion is $ 10^{-10} [m^2\ sec^{-1}]$~\cite{miller2004kinetics} while the surface viscosity $\mu$ is generally {$ 10^{-3}  [kg\ m^{-1}\ sec^{-1}]$}~\cite{durand2006relaxation}. 

This implies that in practice diffusion is much slower than convection, and so we therefore focus here on convective effects.

\subsubsection{Vertex dynamics}

The dynamics of a vertex is calculated through a force balance based on the orientation and tension of each of the three films that meet there. Given the length and surface tension of each of the three segments joining at the vertex, we calculate the vector $ \underline{\gamma}_i = \frac{\underline{l}_i}{{|l_i|}} \gamma_i  $, where $ i = 1,2,3 $. The resultant surface tension force on the vertex is then obtained as the vector sum of the three components, $ \underline{\gamma}_i$, and it is normalized with the sum of the three segment lengths $ l_i $. The normalization allows us to take into account the change in length of segments in the discretization of the films. 

As in the VF model, described below, this net force is balanced by the friction experienced by the vertex and the neighbouring films due to the bounding flat plates which confine the films in our two dimensional system. The friction term is proportional to the vertex velocity, $ \lambda \overline v_{V} $, with drag coefficient $ \lambda $. Then
\begin{equation}
\lambda \underline v_{V} + \frac{\sum{\underline{ \gamma}_i}  } {\sum{l_i}} = 0. 
\label{eq:vertex}
\end{equation}

While the viscous force in Eq.~\eqref{eq:Marangoni} represents the surfactant drag effects opposing the motion of surfactants, this viscous drag force takes into account the contact between films and bounding surfaces. Essentially Eq.~(\ref{eq:vertex}) is an adaptation of the VF model, which we explain in section {\ref{VF}}, to the particular case of the threefold junction.

\subsubsection{Surfactant transfer at the vertex}
\label{Surfactant transfer at the vertex}

Satomi {\it et al.}~\cite{satomi2013modelling} highlight the non-negligible effect of surfactant transfer across the vertex. 
In section \ref{results1} we show (in Fig. \ref{r1a}) that without transfer of surfactant between films the system will reach its final equilibrium configuration far from Plateau's conditions. Therefore, in contrast to the DS model, we assume a continuous exchange of surfactant across vertices.

We use the model introduced by Satomi {\it et al.}~\cite{satomi2013modelling} to relate the amount of surfactant transferred across the vertex to the gradient of surface tension of adjacent films through a coefficient referred as a ``Marangoni" coefficient, $D_m$. The quantity of molecules which moves across the vertex at each step
is proportional to the difference in the surface tension on each neighbouring segment. With the notation of Fig. \ref{fig:s1}, we have 
\begin{equation}
\frac {d(\Gamma_{i} l_i)} {dt} = D_m (\gamma_{i} - \gamma_{k} ).
\label{eq:Satomi}
\end{equation}
where the label $ i $ takes the values 1, 2 and 3 in turn while $k$ takes the values  2, 3 and 1.
Satomi {\it et al.} define $D_m $ to be a function of the vertex size (i.e. liquid fraction), the surface and bulk viscosity and a characteristic system size. They consider a range of values for $ D_m $ between 1 and 5.
Here, we deduce $ D_m $ from dimensional analysis of Eq.~(\ref{eq:Satomi}): we define $ D_m $ to be the ratio between the concentration of surfactant at equilibrium and the surfactant drag coefficient $\mu$. Introducing the length-scale $d_1$, which is the distance between the pins to which the soap films are attached (see Fig. \ref{fig:m1}), we have:
\[
 D_m = \frac{\Gamma_{eq}}{\mu {d_1}}.
\]
By relating $ D_m $ to $ \mu $ we reduce the total number of free parameters in our model.

\subsection{Viscous Froth model + Surfactant transport}
\label{VF}

We now explain how we relax the other strong assumption of the DS model, that of {\em straight} films. 

As explained by Satomi {\it et al.}~\cite{satomi2013diffusion}, the two-dimensional Viscous Froth model (VF model) of Kern {\it et al.}~\cite{kern2004two} is a powerful tool to simulate the rheology of dry foam, in particular to describe the transport of curvature along the film.
The VF model is an extension of the Young-Laplace law. In addition to considering the relation between the gas pressure drop $ \Delta P $ across films, for instance between adjacent bubbles, and the film curvature $ K $, the  VF model introduces a term to describe the local dissipative force opposing the motion of the interface.
The dynamic equation of a single film, in a direction normal to the film, is:
\begin{equation}
\lambda {v_n}^{\alpha} = \Delta P -\gamma K,
\label{eq:VF}
\end{equation}
where $ \lambda $ is the drag coefficient, as in Eq.~(\ref{eq:vertex}), and $v_n$ the normal velocity. The model is simplified by setting $ \alpha = 1 $, even though analysis of quasi-2D foam flow experiments established a range of values for $ \alpha $ between 1/2 and 2/3~\cite{cantat2004dissipation}.
In the present work we consider that all films are subjected to the same pressure on both sides so we neglect the pressure term. Hence we consider the following equation:
\begin{equation}
\lambda {v_n} = -\gamma(\Gamma) K.
\label{eq:VF2}
\end{equation}

As explained by Drenckhan {\it et al.}~\cite{drenckhan2005rheology}, quasi-static models fail to reproduce effects observed in some high velocity experiments of dry foam flowing in channels, while the VF model presents good agreement with experiments. Despite the good agreement, the VF model considers only normal forces acting on the films, while, as confirmed by Cantat {\it et al.}~\cite{cantat2004dissipation}, in the case of higher velocities the presence of both normal and tangential viscous contributions must be included in any useful model.  
We therefore use the VF model to calculate the diffusion of curvature along the films, and instead of considering a constant surface tension, we introduce our calculation of surfactant transport.

\subsection{Dimensionless variables}

A simple dimensional analysis allows us to write a non-dimensional set of equations. We choose $ d_1 $ to be the reference length, which we take to be the distance between the pins, $ d_1 =4 {\rm cm} $ (see Fig. \ref{fig:m1}). Then our dimensionless variables are:
\begin{equation}
\tilde{l} = \frac{l}{d_1}  , \thinspace     \tilde K = K d_1  , 
\label{eq:scaling1}
\end{equation}
\begin{equation}
 \tilde{\gamma}= \frac{\gamma}{\gamma_{eq}}  , \thinspace   \tilde{\Gamma}= \frac{\Gamma}{\Gamma_{eq}} .
\label{eq:scaling2}
\end{equation}

We also define time scales for the three kinematic equations. As illustrated by Kern {\it et al.}~\cite{kern2004two}, the time-scale related to the film relaxation for the VF model, $ T_{\lambda} $, can be defined as 
\begin{equation}
T_{\lambda} = \frac{\lambda {d_1}^2}{\gamma_{eq}}.
\label{eq:TimeLam}
\end{equation}
Dimensional analysis of the vertex dynamics, Eq.~(\ref{eq:vertex}), leads to a time-scale $ T_{V}=T_{\lambda} $, while for our surfactant drag equation, Eq.~(\ref{eq:Marangoni}), we find
\begin{equation}
T_{\mu} =  \frac{\mu {d_1}^2}{\gamma_{eq}}.
\label{eq:TimeMu}
\end{equation}

The three kinematic equations in dimensionless form are:
\begin{equation}
\tilde{v}_n = -\tilde{\gamma} \tilde K \quad \ (VF),
\label{eq:VF_Dless}
\end{equation}
\begin{equation}
\tilde{v}_t  = \frac{1}{\hat{\mu}} \frac {\partial{ \tilde {\gamma}}} {\partial{\tilde s}}  \quad (ST),
\label{eq:Marangoni_Dless}
\end{equation}
\begin{equation}
\tilde{\underline{v}}_{V} + \frac{\sum{ \tilde{\underline{\gamma}}_i}}{\sum{\tilde l_i}} = 0.
\label{eq:vertex_Dless}
\end{equation}

In order to obtain the same time-scale for the three equations, we set $ T_{\mu} = \hat{\mu} T_{\lambda}$, introducing the free parameter $ \hat{\mu} = \frac{\mu}{\lambda} $ in Eq.~(\ref{eq:Marangoni_Dless}). We assume that surfactant drag is weaker than friction with the boundary, choosing a range of values for $ \hat{\mu} $ between $ 0.1 $ and $ 1 $, in section \ref{results1} to investigate the influence of viscous effects on the model's predictions. 

In dimensionless form, the equation of state, Eq.~(\ref{eq:Langmuir}), becomes
\begin{equation}
\tilde {\gamma} = 1 - \hat{E}  \ln \tilde {\Gamma} ,
\label{eq:Langmuir_Dless}
\end{equation} 
leading to our second free parameter  $ \hat{E} = \frac{E}{\gamma_{eq}} $, which is related to the elastic properties of the films, represented by the Gibbs elasticity $ E $. We estimate the order of magnitude of $ \hat{E} $ to be one in the case of SDS~\cite{durand2006relaxation}. 
The initial condition is that the initial surfactant concentration is $\Gamma_{eq}$, and hence the dimensionless surfactant concentration on each segment is $ \tilde{\Gamma}=1 $. This means that all segments initially have surface tension $\tilde{\gamma} = 1$.

Eq.~(\ref{eq:Satomi}) for the surfactant transfer across the vertex becomes, in dimensionless form,
\begin{equation}
\frac {d( \tilde{\Gamma}_{i} \tilde l_i)} {d \tilde t} = \hat{D}_m  (\tilde {\gamma}_{i} - \tilde {\gamma}_{k} ),
\label{eq:Satomi_Dless}
\end{equation}
where $\hat{D}_m = 1 / \hat{\mu}$

To determine the relaxation dynamics of a system of soap films, we must solve Eqns. (\ref{eq:VF_Dless}) -- (\ref{eq:Satomi_Dless}).

\section{Results}

In this section we describe the qualitative behaviour of the VF+ST model. In particular we compare its prediction of the variation of the film length, $ l_1 (t)$, after a T1 topological process with the predictions of the simpler models described above. We will then determine the concentration of surfactant molecules and the consequent evolution of the surface tension in time, varying the free parameters $\hat E$ and $ \hat \mu$.

\subsection{Reference case}

\begin{figure}[!htb]
	\centering
	\vspace*{-0.4cm} 
	\includegraphics[width=0.7\linewidth]{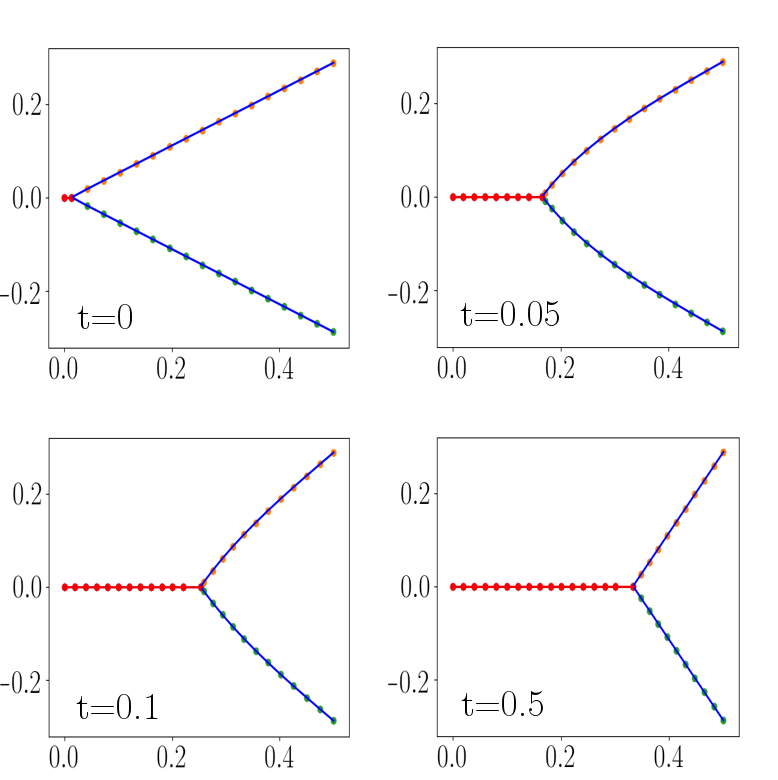}
	\caption{
	Film evolution in the reference case for the VF+ST model. The configuration of the soap films is shown at four different times $t$ during the relaxation, to give an idea of the curvature that is generated in the shrinking films.
	}
	\label{fig:U6_bis}
\end{figure}

\begin{figure}[!htb]
	\centering
	\vspace*{-0.4cm} 
	\includegraphics[width=0.8\linewidth]{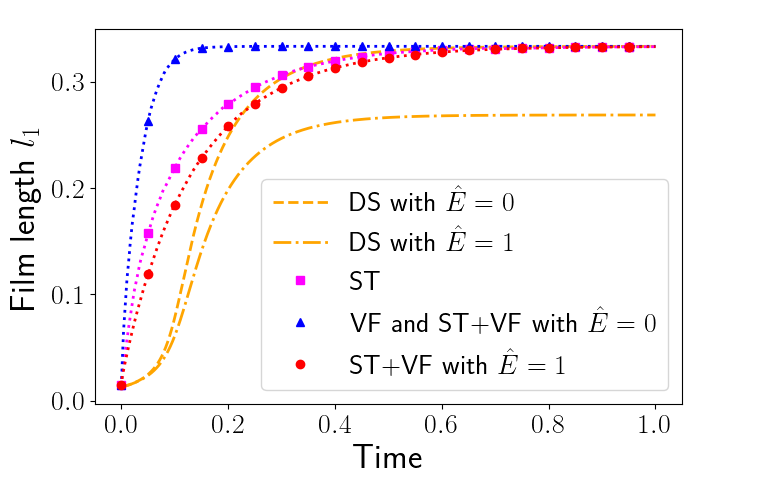}
	\caption{Variation of the film length $ l_1 $ with time, after a T1. We compare the length evolution for the described models, DS, ST, VF and ST+VF. For the DS model we report results for both $ \hat E =0 $ and $ \hat E =1 $. For the ST and ST+VF models both $ \hat E $ and $ \hat \mu $ are one.}
	\label{fig:U6}
\end{figure}

We first summarise, in Fig. \ref{fig:U6}, the different models in terms of their prediction of the film length evolution in time. (i) For the DS model we take $\hat E=0$ or 1 and the factor $\displaystyle \frac{\mu}{\gamma_{eq}}=1 $. (ii) To simulate the pure VF model, we apply Eq.~(\ref{eq:VF_Dless}) and Eq.~(\ref{eq:vertex_Dless}) with all surface tensions equal to one. (iii) For the ST model we solve the system (\ref{eq:Marangoni_Dless}) -- (\ref{eq:Satomi_Dless}) with $ \hat E=1 $ and $ \hat \mu=1 $. (iv) For the VF+ST model we fix $ \hat E=1 $ and $ \hat \mu=1 $ and solve the system (\ref{eq:VF_Dless}) -- (\ref{eq:Satomi_Dless}). The evolution of the films for this case is shown in Fig.\ref{fig:U6_bis}. When $\hat E = 0$ the VF+ST model is effectively reduced to the VF model. 

Grassia {\it et al.}~\cite{grassia2012relaxation} showed that the DS model predicts a slow evolution at short times; this is because it allows slippage between the vertex velocity and the extension rate of surface fluid elements on the film, leading to the trend shown in Fig. \ref{fig:U6}. Moreover, the film reaches its final equilibrium configuration, where the films meet at $ \frac{2 \pi}{3} $, only in the limit of $ E \rightarrow 0 $. 	
This is due to the assumption of constant surfactant concentration on both the shrinking films, $l_2$ and $l_3$ (see Fig.~\ref{fig:s1}). 

In contrast to the DS model, the VF model predicts a rapid rise of the film length towards the final length. Because the force balance at the vertex only depends on the lengths of the segments which join at the vertex, the condition of final equilibrium is quickly realised.

Including the ST model takes into account the variation of the surface tension on each segment and the associated additional viscous effects. For this reason the balance at the vertex is reached more slowly, and consequently the film length evolves more slowly.
  
Not surprisingly the combination of these last two models, VF+ST, leads to an even slower length evolution.

\subsection{The effects of surface viscosity with and without surfactant transfer across the vertex}
\label{results1}

In order to evaluate the effect of  surface viscous components of our model we fix the elastic parameter $\hat{E}=1$ and we vary $ \hat{\mu} $ between 0.1 and 1. 

\subsubsection{Without surfactant transfer  across the vertex}

As a first step we consider how the system of soap films relax without surfactant transfer between films (across the vertex), i.e. when $\hat D_m$ is decoupled from $\hat{\mu}$ and then set to zero.

\begin{figure}[!htb]
	\centering	
	\subfigure[\label{r1a}]{\includegraphics[width=.8\linewidth,valign=t]{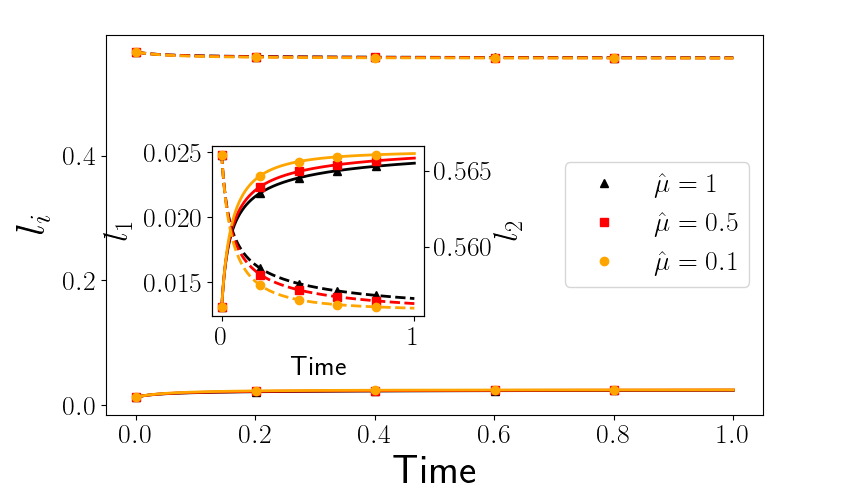}}
	\subfigure[\label{r1b}]{\includegraphics[width=.8\linewidth,valign=t]{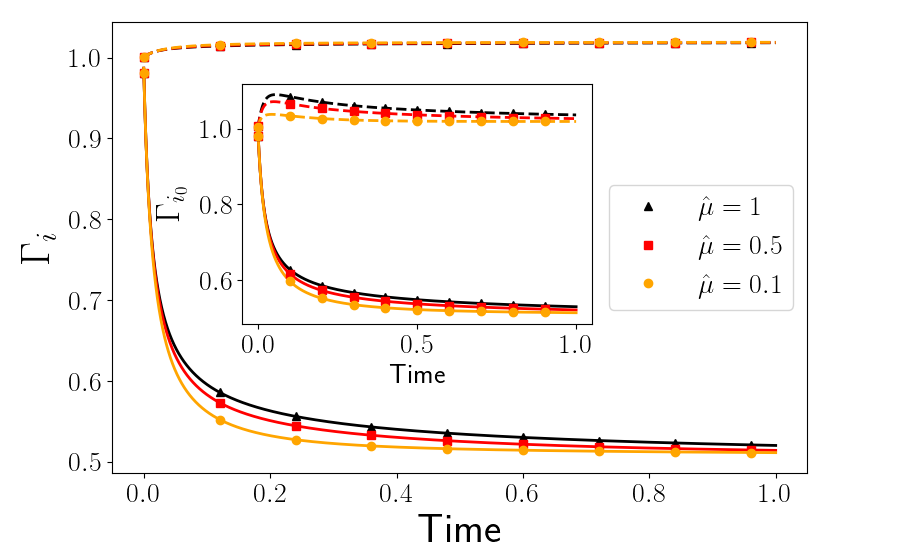}}
	\subfigure[\label{r1c}]{\includegraphics[width=.8\linewidth,valign=t]{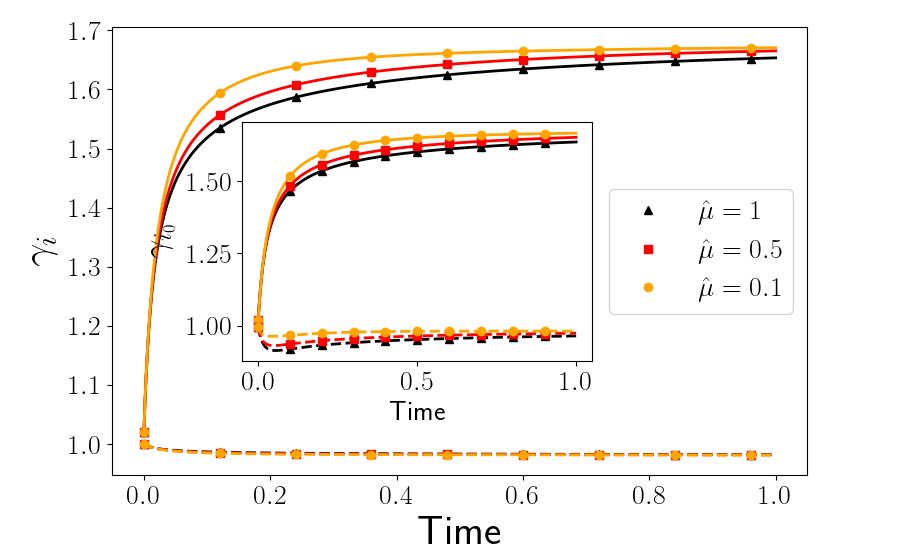}}
  \caption{ST+VF model results for different $\hat{\mu}$, with $\hat E = 1$ and $D_m = 0$, for the stretching film $l_1$ (solid lines) and the shrinking film $l_2$ (dashed lines). (a) Variation of film lengths. (b) Variation of average surfactant concentration $\Gamma_i$. (c) Variation of average surface tension $\gamma_i$. In (b) and (c) the insets show the same data for the segments which meet at the vertex, $\Gamma_{i_0}$ and $\gamma_{i_0}$.} 
	\label{fig:r1}
\end{figure}

Fig.~\ref{r1a} shows how, following a T1, the newly-created film $l_1$ starts to grow while the lateral film $l_2$ shrinks. Consequently, the surfactant concentration on the stretching film $l_1$ decreases while the concentration on the shrinking film $l_2$ increases. This is shown in Fig. \ref{r1b} where we plot the average surfactant concentration on both films. Fig. \ref{r1a} also shows, for each film, the value of $\Gamma$ on the segments which meet at the vertex. The concentration $\Gamma_{2_0}$  increases, due to the reduction of the film length, and then decreases as a consequence of the homogenisation of segment lengths within the same film. 

As expected from Eq.~(\ref{eq:Langmuir_Dless}), the opposite trend is found for the surface tension: $\gamma$ increases on the stretching film $l_1$ and decreases on the shrinking film $l_2$, as shown in Fig. \ref{r1c}. 

Moreover, Fig. \ref{fig:r1} shows how the viscous parameter $\hat{\mu}$ affects the film evolution. In particular, increasing the value of $ \hat{\mu} $ slows down the relaxation, reducing the variation in film lengths, because the surfactant concentration is higher, and the tension consequently lower, close to the vertex. 

Due to the condition $\hat D_m=0$, suppressing surfactant transfer across the vertex, the relaxation of the film stops when the angles between the films are still far from $2\pi/3$, Fig.~\ref{r1a} shows the remarkable lack of stretch of both films. This is caused by our assumption for the starting surfactant concentration value. We assume all films having a starting equilibrium concentration, $\Gamma_{eq}$. As a consequence the forces acting on the vertex are suddenly balanced, but the well-known Plateau's laws are not satisfied.
Because the T1 is an instantaneous process is not straightforward to estimate a coherent value for the surfactant concentration of the new created short film. We decide to assume $\Gamma_{eq}$ as starting value rather than fix an arbitrary value bigger than $\Gamma_{eq}$ for the new film. In the next section we show how our assumption is physically meaningful if we allow mass transfer between adjacent films. The condition $\hat D_m \neq 0$ is therefore necessary in order to use our surfactant transfer model as a predictive tool.

\subsubsection{With surfactant transfer across the vertex}

\begin{figure}[!htb]
	\centering	
	\subfigure[\label{r2a}]{\includegraphics[width=.8\linewidth,valign=t]{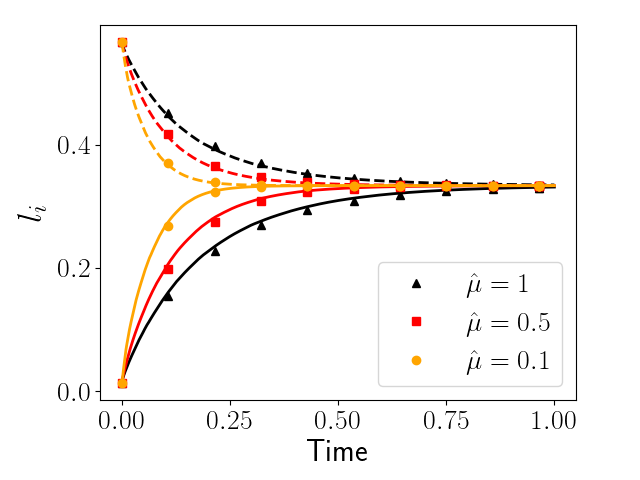}}
	\subfigure[\label{r2b}]{\includegraphics[width=.8\linewidth,valign=t]{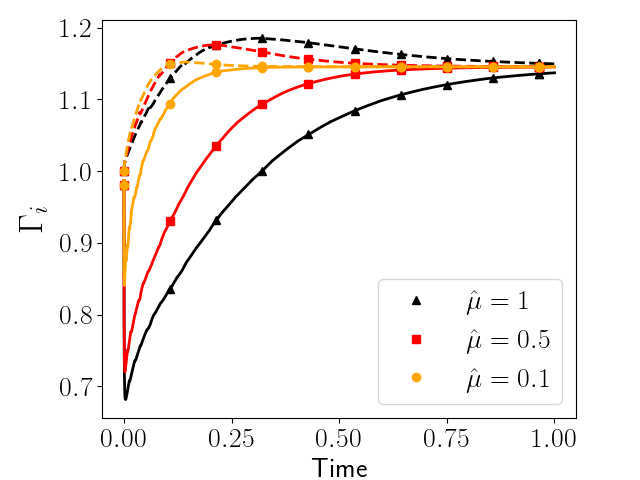}}
	\subfigure[\label{r2c}]{\includegraphics[width=.8\linewidth,valign=t]{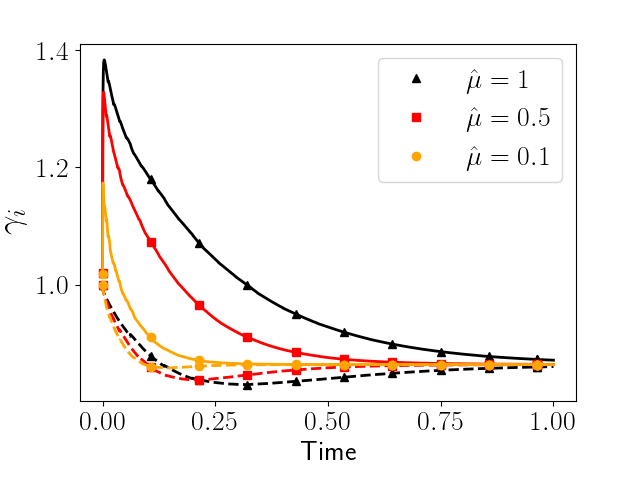}}
  \caption{ST+VF model results with surfactant transfer across the vertex, with coefficient $\hat{D}_m  = 1 / \hat{\mu} $. Notation and other simulation details as for Fig.~\ref{fig:r1}.}
\label{fig:r2}
\end{figure}

To allow full relaxation to an equilibrium that obeys Plateau's rules we now relax this condition, and introduce surfactant transfer across the vertex. Fig. \ref{fig:r2} shows the results when $\hat{D}_m  = 1 / \hat{\mu}$. 
Increasing $\hat{\mu}$ again decreases the rate of the length variation, shown in Fig. \ref{r2a}, because the lower surface tensions at the vertex takes more time to be balanced.

In Fig. \ref{r2b} we observe the influence of surfactant transfer across the vertex on surfactant concentration. As before, the initial trend is that the surfactant concentration on the newly-created film $ l_1 $ decreases as a consequence of it stretching. But this trend is soon reversed because of surfactant transfer, across the vertex, from the shrinking film, on which surfactant is building up. The relaxation then continues uniformly until the final equilibrium surfactant distribution is reached.

Moreover, since $\hat{D}_m $ influences {the rate at which} surfactant molecules move across the vertex, through Eq.~(\ref{eq:Satomi_Dless}), we observe how the surfactant transfer and therefore the variations in surfactant concentration occurs more quickly with small values of $\hat{\mu}$. On the other hand, on the shrinking film $ l_2 $ the surfactant transfer reduces variations in $ \Gamma $. The consequent variations in surface tension $\gamma$ (Fig. \ref{r2c}) are also smaller, so that the final equilibrium is reached more quickly at low  $\hat{\mu}$.

As expected from the observation that soap films minimize their length~\cite{cantat2013foams}, the total length at the final time is less than the total initial length. The final value of the surfactant concentration is therefore higher than the initial (unit) value, as shown in Fig. \ref{r2b}. 

\subsection{The effect of surface elasticity}

\begin{figure}[!htb]
	\centering	
	\subfigure[\label{r3a}]{\includegraphics[width=.8\linewidth,valign=t]{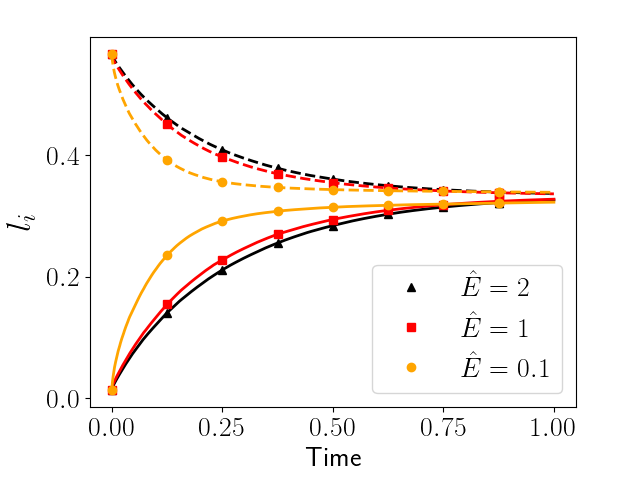}}
	\subfigure[\label{r3b}]{\includegraphics[width=.8\linewidth,valign=t]{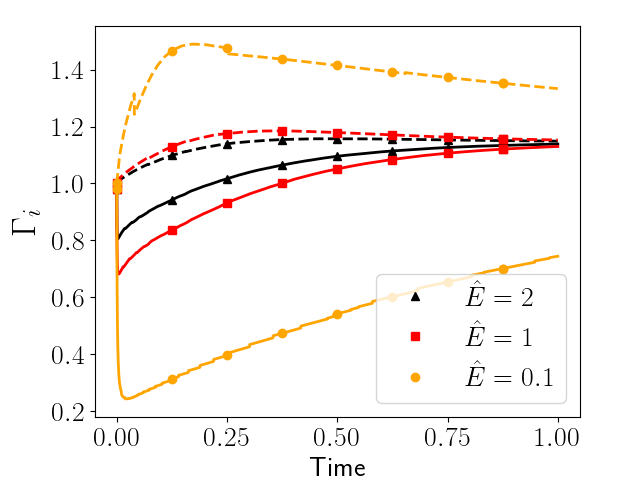}}
	\subfigure[\label{r3c}]{\includegraphics[width=.8\linewidth,valign=t]{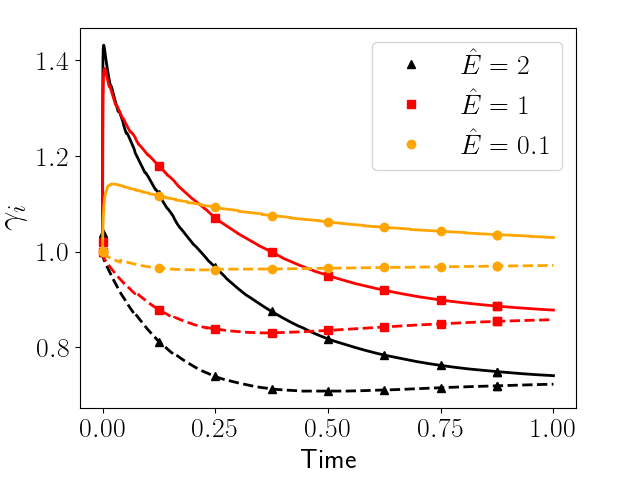}}
  \caption{ST+VF model results with varying elasticity $\hat{E}$. Here $\hat{\mu} = \hat{D}_m = 1$. Notation and other simulation details as for Fig.~\ref{fig:r1}.}
\label{fig:r3}
\end{figure}

We now turn to the effect of the surface elasticity parameter, $\hat{E}$, on the evolution of the soap films. We fix the surface viscosity at $\hat{\mu} =1 $ and hence the coefficient $\hat{D}_m $ is constant. We vary $\hat{E}$ between $0.1$ and $2$, which affects the surface tension through the Langmuir equation of state, Eq.~(\ref{eq:Langmuir_Dless}). With larger values of $\hat{E}$, the surface tensions acting at the vertex are weaker, and therefore the film takes longer to reach its final length, as shown in Fig. \ref{r3a}.

In Figures \ref{r3b} and \ref{r3c} we report the average values of surfactant concentration and surface tension for the two films. The initial concentration on each film is one, but now the initial surface tension depends on $\hat{E}$ according to the Langmuir equation of state, Eq.~(\ref{eq:Langmuir_Dless}).
Similarly, the final value of surface tension, while equal on the two films, also depends on $\hat{E}$. 
Fig. \ref{r3c} shows that the surface tension $\gamma_i$ at equilibrium is inversely proportional to $\hat E$.

\begin{figure}[!htbp]
	\centering
	\includegraphics[width=0.8\linewidth]{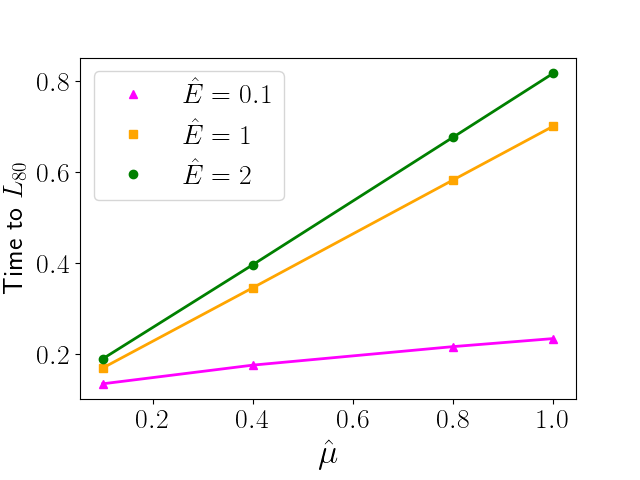}
	\caption{Variation of the time to reach $80 \%$ of the final length for the newly-created film $l_1$, as a function of $\hat E$ and $\hat{\mu}$.}
	\label{fig:E1}
\end{figure}

To summarise our results we calculate the time for the stretching film to reach $80 \%$ of its final length during the relaxation process. Fig. \ref{fig:E1} shows that increasing either the viscous effects ($\hat{\mu}$) or the surface elasticity ($\hat E$) results in an increase in the time required for the film to relax to equilibrium. For each value of $\hat E$, this relaxation time increases linearly with $\hat{\mu}$.

\subsection{Experimental data fitting}

\subsubsection{Experimental details}

The experimental setup is depicted in Fig.~\ref{fig:m1}. The device consists of two flat Plexiglass plates held a distance of $ 2$cm apart by four metal pins. Two pins are fixed towards one end of the plates, a distance $ d_1 = 4$cm apart. The other two pins are attached to a sliding rod which allows the distance $ d_2 $ from the fixed pins to be varied, as shown in Fig. \ref{fig:m1}, up to about $10$ cm. The device is immersed in a solution of water and surfactant in order to create five films of the foaming solution between the pins. Moving the rod causes rearrangements of the films, and when pushed sufficiently the rod induces a T1 process.

Image analysis of a video of the experiment was used to extract the length evolution of the film $ l_1 $ after the T1~\cite{alexia}. The film length is then normalized by its final length and plotted against time in seconds in Fig. \ref{fig:final3}.
In order to study the influence of the rheological interfacial properties on the T1 process we use two different foaming agents: Sodium dodecyl sulfate (SDS), at a concentration of $ 4.80 g/l $, and $\beta$-Lactoglobulin (BLG), at a concentration of $ 50 g/l $ and $ pH = 7 $.
For the equilibrium surface tensions we take the values $ 38 mN/m$ (SDS)~\cite{durand2006relaxation} and  $ 48 mN/m$ (BLG)~\cite{alexia}.

\subsubsection{Validation of the VF+ST model}

We validate the VF+ST model by fitting the experimental data for film length evolution after a T1 for each of these foaming solutions. The numerical model is used to simulate the film evolution, then a fitting procedure is used to extract a prediction of the material parameters. 

\begin{figure}[!htb]
	\centering
	\includegraphics[width=0.4\textwidth]{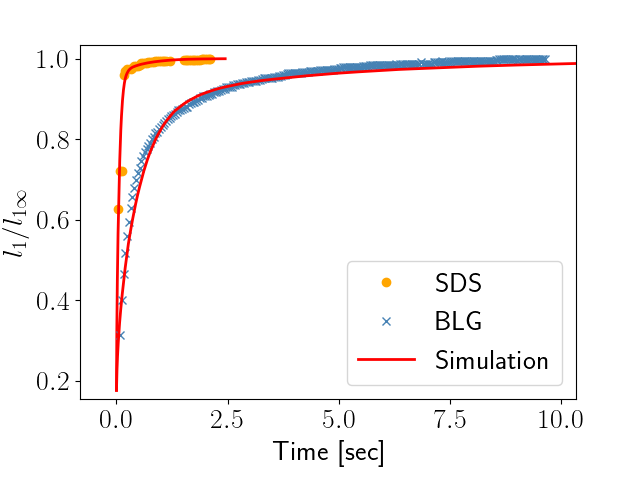}
	\caption{Experimental data for the length of the newly-created film $ l_1 $ after a T1 with surfactant solutions of SDS and BLG is compared with a simulation with the VF+ST model which has been optimized by fitting values for the parameters $\hat{\mu}$ and $ \hat{E}$.
	}
	\label{fig:final3}
\end{figure}

As shown in Fig.~\ref{fig:final3}, the films made with SDS reach their final length much faster than with BLG. We first scale our numerical simulation results by choosing the time-scale $T_{\lambda}$ to be the time at which the film length is $85 \%$ (BLG) or $99 \%$ (SDS) of its final equilibrium value. The different percentages are chosen to account for the different rates of relaxation, ensuring that for the faster evolution there are enough data points available for a good fit.
 
Once we know $ T_{\lambda} $ we calculate $ \lambda [kg \ m^{-1} sec^{-1}]$ through the definition $T_{\lambda}= \displaystyle \frac{\lambda {d_1}^2}{\gamma_{eq}}$ (Eq.~(\ref{eq:TimeLam})), taking the distance between the pins, as the reference length, $ {d_1}=4$cm, and using the known equilibrium surface tension for SDS and BLG. The values obtained for the two samples are reported in Table~\ref{tab:tab3}. 

We then convert the simulation data to real times and implement a non-linear least-squares fit of our model to the experimental data with parameters $ \hat{\mu} $ and $ \hat{E} $. 
We impose a relative tolerance of $ 10^{-3} $ for the calculation.
In Fig. \ref{fig:final3} we show the fit of the simulation for the normalized film length against time. In the case of BLG the fit doesn't exactly capture the early time evolution while at long times it reaches the final value slightly slower than the experimental data, causing an overestimation of the surface viscosity. 

Finally, given the optimized model parameters $ \hat{\mu} $ and $ \hat{E} $, we deduce $ \mu = \lambda \hat{\mu}[mPa \ m \ sec] $ and $ E = \gamma_{eq} \hat{E} [mN/m]$. The values predicted for both the SDS and BLG solutions are reported in Table~\ref{tab:tab3}; we discuss each prediction in the following sections.
\begin{table}
	\caption{Drag coefficient, $ \lambda $, surface viscosity, $ \mu $ and Gibbs elasticity, $ E $ predicted by a fit of the VF+ST model to experimental data for solutions of SDS and BLG.} \label{tab:tab3}
	\begin{tabular}{ |c|c|c|c| } 
		\hline
		&  $ \lambda $ &	$ \mu   $ & $E $ \\ 		
		&  $   [Pa \ sec]$ &	$   [mPa \ m \ sec]$ & $ [mN/m]$ \\ 
		SDS &	$0.74 \pm 0.02$ &$ 9.96 \pm 0.02$ & $42.8 \pm 0.02$\\		
		BLG &	$3.80 \pm 0.02$ &$212 \pm 0.02$ & $156 \pm 0.02$\\	
		\hline		 
	\end{tabular}
\end{table}

\subsubsection{Predicting the drag coefficient $ \lambda $}

The values obtained for the drag coefficient $ \lambda $ are reported in Table~\ref{tab:tab3}:
we find that $\lambda$ is about six times greater for BLG compared to SDS, suggesting that $\lambda$ is not a property just of the plates, but also of the solution.

Drenckhan {\it et al.}~\cite{drenckhan2005rheology} used the VF model to simulate two-dimensional flowing foams and they predict a value for the drag coefficient $ \lambda $ which is slightly higher than our prediction. Their experimental setting was different, so we don't necessarily expect to find exactly the same value for $\lambda$. However, we repeated our fitting procedure for our experimental data using the VF model without surfactant transfer and, in agreement with Drenckhan's results, we find that the VF model on its own predicts a higher value for the drag coefficient: $ \lambda_{SDS} = 1 Pa \ sec $ and  $ \lambda_{BLG} = 11 Pa \ sec $. In general, we suggest that applying the VF model on its own overestimates the drag coefficient.

Moreover, using the VF model without surfactant transfer, we notice that although at very short times the VF model is able to fit the data for anionic surfactants, at long times the introduction of additional surface viscous factors within the VF+ST model is crucial in order to fit data for both anionic surfactants and proteins.
 
\subsubsection{Predicting the surface viscosity $ \mu$}

There are two contribution to the surface viscosity, which corresponds to $ \mu $ in our model, shear and dilatation.
Edwards {\it et al.}~\cite{edwards1991interfacial} explain why the experimental measurement of the interfacial dilatational viscosity, $ k $, is more challenging than the measurement of the interfacial shear viscosity, $ \mu_s $~\cite{edwards1991interfacial}. This is due to the coupling that arises between interfacial dilatational viscous and elastic effects. They also show how both components of the complex dilatational modulus, $ E^\prime$ and $ E^{\prime\prime}$ can be calculated as a function of the Gibbs elasticity $E$, a diffusion parameter $ \tau $ and the frequency of oscillation during measurements $ \omega $. 

In the case of an insoluble layer, when $ \tau\rightarrow 0 $, $E^\prime$ is reduced to the Gibbs elasticity while the ratio $  E^{\prime\prime}/\omega $ goes to zero~\cite{edwards1991interfacial}, and then the sum of the surface viscosities $ \mu_s +k $, which corresponds to our $ \mu $, may be determined experimentally.
Despite these considerations, it is not straightforward to find surface viscosity values in the literature. Therefore for the SDS anionic surfactant we compare our prediction for $ \mu $, in Table~\ref{tab:tab3},  with the DS model prediction~\cite{durand2006relaxation}. They calculate $ \mu_{SDS} \simeq 1.3  [mPa \ m \ sec]$ while we found $ \mu_{SDS} \simeq 10  [mPa \ m \ sec]$. Despite the large possible range of variation in the values, our prediction is rather low.

There is even less published data for BLG, due to the difficulties mentioned above, and it is not simple to find surface viscosity values in literature. This underlines the importance of having a tool such as the VF+ST model to predict such quantities. Nevertheless, we observe that our model probably overestimates the surface viscosity values, reported in Table~\ref{tab:tab3}, especially in the case of proteins. This suggest the possibly that there are still missing characteristics which have to be considered in order to model foaming solutions.

\subsubsection{Predicting the Gibbs elasticity $ E$}

Finally we discuss our prediction for the Gibbs elasticity. We find a good agreement with the results of Durand and Stone for SDS. Their model predicts $ E = 32  mN/m $~\cite{durand2006relaxation}, while we find $ E = 42.8  mN/m $.

Recall that, for an insoluble layer, the real component of the dilatational surface modulus, $ E^\prime $, is equal to the Gibbs elasticity. In order to evaluate the accuracy of our prediction in the case of BLG, we seek characteristic values for its dilatational surface modulus.  

The surface dilatational modulus of protein solutions is strongly dependent on the concentration and pH of the sample~\cite{peng2017foams}. The BLG sample used in our experiments has a concentration of $ 50 g/l $ and $ pH = 7 $, for which the real component of the dilatational surface modulus is $ E^\prime \approx 40 $~\cite{alexia}, but values up to 110 have been found \cite{ulaganathan2017beta}. Comparing these values with our prediction, reported in Table~\ref{tab:tab3}, we conclude that our prediction of the Gibbs elasticity has a reasonable order of magnitude.

\section{Conclusion}

Two-dimensional dry foams have been widely studied and simulated, yet analysis of the viscous effects present in non quasi-static simulations still requires further investigation.

Here we analyse the advantages and disadvantages of existing models. We merge different ingredients to create a model which simulates the film evolution taking into account surfactant transport. In particular we look at the film relaxation after a T1 rearrangement. We consider the variation of surfactant concentration, and therefore the variation of surface tension, along both stretching and shrinking films. As a consequence we introduce an additional dissipative term which represents the viscous effects deriving from gradients in the surface tension and we allow surfactant transfer between adjacent films,
Our model is characterized by, in addition to the time scale $ T_{\lambda} $, two free parameters $ \hat{E} $ and $ \hat{\mu} $, related to elastic and viscous effects respectively.
We investigate the impact of these two characteristic parameters on film evolution and estimate the flow of surfactant molecules along and between the films as a function of both $ \hat{\mu} $ and the surface tension gradient.

We validate our VF+ST model by fitting experimental data. In particular we look at foaming solutions with anionic surfactants and proteins. The VF+ST model is able to describe the film length evolution for both solutions. We apply the scaling factor $ T_{\lambda} $ to the simulation results in order to match the experimental data and calculate the drag coefficient $\lambda$ for both samples. Then we conduct a model optimization in order to find the best values for the parameters $ \hat{E} $ and $ \hat{\mu}$. Finally we predict relevant parameters, such as Gibbs elasticity and surface viscosity, for the experimental data.   

The VF+ST model can now be applied to ordered (hexagonal) and disordered foams, with many interacting bubbles, subjected to different strains.

Despite the promise of our model, a further extension may be the inclusion of a model for soluble interfaces. We could also introduce and analyse the action of the disjoining pressure within the film, and hence extend our work to the case of wet foams.

The film characterization proposed here can be extended to the foam scale. As Cantat {\it et al.}~\cite{cantat2004dissipation} highlight, in a fast-flowing foam the viscous dissipation due to the tangential velocity of surfactant molecules is not negligible. Our model can be used to investigate foams flowing at high velocities.

\section*{Acknowledgments}
The authors gratefully acknowledge funding from the MCSA-RISE project MATRIXASSAY (ID: 644175) and the UK Engineering and Physical Sciences Research Council (EP/N002326/1).

\end{document}